# Decoding the Meaning of Success on Digital Labor Platforms: Worker-Centered Perspectives

Pyeonghwa Kim, Syracuse University, USA
Charis Asante-Agyei, Syracuse University, USA
Isabel Munoz, Syracuse University, USA
Michael Dunn, Skidmore College, USA
Steve Sawyer, Syracuse University, USA

What does work and career success mean for those who secure their work using digital labor platforms? Traditional research on success predominantly relies on organizationally-centric benchmarks, such as promotions and income. These measures provide limited insights into the evolving nature of work and careers shaped at the intersection of digital labor platform technologies and workers' evolving perspectives. Drawing on data from a longitudinal study of 108 digital labor platform workers on Upwork, we (1) identify seven dimensions of success indicators that reflect workers' definitions of success in platform-mediated work and careers, (2) delineate three dimensions of digital labor platforms mediating workers' experiences of success and (3) examine the shifting perspectives of these workers relative to success. Based on these findings, we discuss the implications of platform-mediated success in workers' labor experiences, marked by platformic management, standardization, precarity and ongoing evolution. Our discussion intertwines CSCW scholarship with career studies, advancing a more nuanced understanding of the evolving perspectives on success in platform-mediated work and careers.

## 1 INTRODUCTION

We focus attention on concepts of career and success in the context of workers seeking work on digital labor platforms. We do so because digital labor platforms mediate these workers' labor experiences and occupational perspectives, such that platform design is bound up in worker's career plans [13,23,73,74,115].

To do this we focus on workers who make a living outside of traditional organizations. Often called freelancers, gig workers, or independent contractors, these workers made up 12% of the global labor force and 38% of the U.S. workforce in 2023 [140,149]. Different from the more commonly studied work of rideshare and food delivery workers, online freelancers pursue jobs that align with their educational backgrounds, occupational interests and career aspirations.Many of these workers are increasingly pursuing work and careers on digital labor platforms like Upwork.com, which have 163 million registered users and show consistent yearly growth [54,84,130].

The analysis we report on below shows there are at least seven dimensions of subjective success indicators in digital platform-based work. Building from this, we advance three contributions to the Computer-Supported Cooperative Work (CSCW) and cognate communities. First, we make explicit that worker's perceptions of success are bound up in the design of the platform [13,121]. Second, we contribute to our understanding of the roles of platforms in shaping career success [79]. Third, the conceptual developments outlined here provide a bridge between contemporary CSCW and careers scholarship.

### 1.1 Intersecting Discourses on the Changing Nature of Work and Careers

Like other CSCW researchers, we see the need for greater scholarly attention to the long-term career trajectories and technology-enabled support necessary for successful career development among digital labor platform workers [5,23,70,73]. One of the prerequisites for these efforts is to understand how these workers interpret success on digital labor platforms. Indeed, Seibert and colleagues' (2024) literature review concludes that research on career success among workers in non-standard work arrangements is a crucial, yet significantly understudied area [121].

Motivated by this awareness, the research reported on here is framed by three intersecting discourses on the changing nature of work and careers: (1) the rise of non-traditional career trends, coupled with shifting orientations toward career success, (2) the emergence of platform-mediated careers and (3) growing scholarly attention to platform-mediated careers in the CSCW and cognate communities. We define career success as the positive material and psychological achievements resulting from work experiences [96,121,124]. 'Work experiences' are traditionally gained from a position or occupation within organizational boundaries [46,128]. This means career success is primarily viewed through organization-centric indicators such as corporate rank, salary and colleague relations [55,80,116].

In contrast, beginning in the 1980s, career trends began to shift away from the conventional linear career models that emphasized organizational influence over career success to contemporary multidirectional models that highlight individuals' proactiveness in carving out the meaning of career success [12,19,33,63,82]. Within this broadened view of a career, work experiences are acquired across multiple organizational, occupational and geographical and circumstantial settings. This shift was theorized through a number of alternative career models, to include boundaryless [8,9], protean [62,64], kaleidoscope [98,127] and portfolio approaches [35,65,99]. The nature of success shared in these models often reflects individuals' intrinsic values, needs and aspirations [9,25,55,59,133].

From the late 1990s, the emergence of digital labor platforms has led to a steadily increasing number of individuals seeking work through these platforms [37,74,130,142]. Digital labor platforms (also referred to as online labor platforms or gig work platforms) are overarching terms for online entities that connect supply and demand in the labor market [76, 107]. Digital labor platforms such as Upwork and Fiverr attract workers pursuing self-defined success standards such as spatio-temporal flexibility, skill development, economic rewards and passion [4,137]. And, this phenomenon is expected to shape an emerging career trend. For instance, [40] found that more than half of platform workers across six global regions are under the age of 30, and they are young professionals starting their careers on these platforms. Incoming cohorts of workers, such as Generation Z (those born between 1995 and 2010), prefer to pursue their careers on digital labor platforms [17,53,111,152].

From the late 2010s, the confluence of these trends began to be visible in the literature of CSCW and adjacent communities [86,100]. These authors noted that the design of the platforms is increasingly mediating the structure of work and the career possibilities of workers. That is, the rise of online work through digital labor platforms draws together scholars of work and labor and scholars of CSCW.

Researching career success benefits workers, organizations and - increasingly - platform owners. On an individual level, workers need to design effective career strategies to achieve their career goals [93]. While not a focus of the work reported here, at the organizational level, career success research informs the designing of career advancement structures, thereby improving worker satisfaction and meaningfulness at work [26,123,126]. As digital labor platforms continue to evolve and - in doing so to - shape workers' career trajectories, we examine what it means to be successful in this emerging career path.

### 1.2 Research Questions

The rise of non-standard work arrangements, the changes in career paths, the emergence of digital labor platforms - and their role in framing work and possible career success - lead us to pursue three research questions (RQ):

**RQ1.** How do digital platform workers interpret career success?
**RQ2.** How do digital labor platforms shape these workers' interpretation of career success?
**RQ3.** How do these workers' interpretation of career success evolve over time?

As discussed below, to answer these RQs we draw from a multiyear longitudinal approach to the study of 108 digital labor platform workers on Upwork, one of the world's largest digital labor platforms [84, 130]. Using data from four rounds of interviews, surveys and profile archival data collection, we examine how digital labor platform workers' perspectives on success are constructed and evolved throughout their participation in the digital labor market.

The rest of the paper is structured as follows. In the next section we discuss the existing literature on digital labor platforms and career success. Then, we present our method, data, and analytical approach. This is followed by our findings, which set up our discussion of the implications of this study.

## 2 RELATED WORK

This review draws from relevant literature from multiple scholarly spaces that bears on digital labor platforms, online freelancing, careers and career success. Like others, we see the value of using the concepts and insights from organizational studies to advance CSCW scholarship [18] This advances CSCW in two ways: (1) It provides a structured means to understand career success on digital labor platforms, serving as a conceptual lens to unify previously fragmented findings in the CSCW literature regarding platform workers' careers and their long-term desires and aspirations [25,115]. (2) This lays the foundation for future research and design practices on computer-supported career development, addressing a nascent area of scholarly effort within CSCW [73].

### 2.1 Digital Labor Platforms and Online Freelancing

Digital labor platforms are market-makers in the platform economy, serving as intermediaries and in doing so creating a two-sided market between supply and demand [56]. The supply encompasses individuals registered on the platform, offering labor in exchange for compensation. The demand is constituted by individuals or companies that pay for work [75,91]. Although classifications differ, the platform work literature typically distinguishes among online freelancing, crowdwork, and delivery and transportation platforms based on contract type, degree of time/location flexibility, skill requirements and client interaction [21,27,81,87,90,108,118].

The CSCW and cognate communities have paid increasing attention to online freelancing platforms in the last five years, driven by their growing presence in the labor market and the comparatively limited scholarly exploration in comparison to other types of platform work [7,30,40,54,86]. Online freelancing is characterized by (1) longer-term contracted projects, (2) a greater degree of spatio-temporal flexibility, (3) intensive knowledge work and advanced skill requirement and (4) continued interaction and collaboration with clients [23,45,87,109,146]. Online freelancing platforms operate in a dual capacity as an ostensibly 'neutral intermediary', brokering labor transactions between supply and demand [57]. Simultaneously, these platforms function as a 'shadow employer', exerting inadvertent or intentional influence over workers' daily labor processes and long-term career trajectories [6,23,27,31,51,79,87,103,129].

This arrangement makes it hard to apply most contemporary models of career and career success. This is because career and career success models are rooted in concepts of traditional organizational settings and alternative employment structures - arrangements that do not rely on digital labor platforms [29,49,68,113,121,133,134]. These models of career and success come from labor and organizational studies and there is now a growing awareness of these in CSCW [22,23,29,31,115].

For example, Blaising and colleagues (2021) investigate long-term experiences of workers as they navigate careers on an online freelancing platform. Their findings reveal that the platform-mediated career serves as a space for testing, confirming and developing multiple career potentials, despite challenges such as financial and emotional burdens [23]. [115] examine career transition among workers on a crowdwork platform, in search of greater employment stability and professional fulfillment.

Despite Blaising and colleagues' (2021) exploratory research highlighting the challenges faced by workers in navigating platform-based careers, there is much to learn [23]. Recognizing this, Ashford and colleagues (2018) calls for a qualitative study that delves into how individuals in the non-standard work arrangements define success [13]. They find that traditional metrics of success may not adequately capture the diverse and evolving criteria that shape the career aspirations and priorities of those in non-traditional career paths. This motivates the need for additional scholarly exploration to better understand what success means in these arrangements.

## 2.2 Careers and Success

Like others, we define a career as "the unfolding sequence of a person's work experiences over time" [10:8]. Career success is a measure that captures both reflections on past achievements and projections toward future goals [52]. Career success has been operationalized through both objective (or extrinsic) and subjective (or intrinsic) lenses [1,20,55,71,72,106,131].

Objective career success are measurable and verifiable achievements in an individual's work-related endeavors, while subjective career success involves an individual's personal assessment of their career based on dimensions that hold significance for them [16,71,124,143]. Objective career success is often represented by standardized benchmarks that can be compared across individuals such as salary and promotions [67,136]. Subjective career success is operationalized in a more holistic manner such as satisfaction and overall career perception. [58,138].

Objective and subjective career success models are further contextualized in a broader societal and economic contexts, undergoing a transition from traditional to contemporary era [11,33,36]. Objective career success has been mostly understood within an organizationally-centric paradigm [8,61,62,150]. Career success indicators are often hierarchically constructed, centrally planned and controlled, and formally documented [33].

Subjective career success has gained attention more recently, given the trajectory of employment relations that are more contractual, project-based and contingent [33]. The idea of organizations providing employment security and career training have diminished. Instead, there is heightened significance placed on individual workers, their responsibilities, and their subjective perceptions regarding career success [8,61,62,150].

With the emergence of platform work, subjective career success holds greater significance. Unlike traditional career structures, platform-mediated work often lacks conventional benchmarks of success, such as opportunities for promotion or standardized increases in compensation. In this context, intrinsic markers may hold greater significance as a compass to navigate the ever-evolving platform labor environments. Identifying intrinsic measures of success will help promote diverse and inclusive values for workers while informing the (re)design of the platform work experience to support the achievements of workers in their professional journeys.

### 2.3 Twelve Dimensions of Career Success

As a theoretical framework guiding our analysis of the dataset, we employ Seibert and colleagues' dimensions of subjective career success building [121] (see Table 1).

Drawing on a review of career success literature published between 1986 and 2021, Seibert and colleagues (2024) present eight conceptualizations of subjective career success: (1) financial concerns, (2) career advancement, (3) interpersonal relations, (4) challenge and mastery, (5) meaning and impact, (6) self-development, (7) career control and (8) work-life interface [11]. We further note that "dimension" denotes a high-level aspect of subjective career success, while "indicator" signifies one of the measures that collectively form the higher-level dimension. Each career success dimension includes multiple indicators, each highlighting different facets of the overarching dimension.

We complemented this framework after reviewing relevant publications across multiple literatures in order to identify and understand the nuanced dimensions of subjective career success. This effort led us to add four more dimensions: (9) performance, (10) recognition, (11) fulfillment and (12) agency. The framework serves as a guiding structure for our data analysis.

Table 1. Subjective career success dimensions and indicators

| Dimensions | Pertinent indicators |
|---|---|
| Financial* | · Income [55,58]*<br>· Financial success [122]*<br>· Financial security [26]* |
| Career advancement* | · Career advancement [55,58]*<br>· Social status [122]* |
| Interpersonal relations* | · Colleagues relations [55]*<br>· Stable relationship with partner and children [47] |
| Challenge and mastery* | · Learning, growing and being challenged [94]<br>· Winning/overcoming challenges [145] |
| Meaning and impact* | · Perceived contributions [42]*<br>· Positive impacts [26]*<br>· Contributing to a community [48] |
| Self-development* | · Skill learning opportunity [55]*<br>· Developing new skills [58]* |
| Career control* | · Employability [122]*<br>· Desirable job opportunities [122]* |
| Work-life interface* | · Work-life balance [26]*<br>· Happiness from private life [55]*<br>· Having a life outside work [94] |
| Performance | · Job performance [94]<br>· Sense of achievement [132] |
| Recognition | · Appreciation of achievement by others [42]<br>· Reputation and fame [145] |
| Fulfillment | · Happiness from job [147]<br>· Intrinsic fulfillment from work [151]<br>· Fun and enjoyment/doing interesting things [94] |
| Agency | · Work freedom [47]<br>· Independence and liberty [145] |

* Original dimensions and indicators in Seibert and colleagues' (2024) work [121]

## 3 METHODS

We present our research approach, beginning with a description of the research site, Upwork[1]. Following this, we describe our longitudinal data collection process and the multiple data collection techniques. Finally, we detail our thematic analysis methods used to derive the insights presented in the findings section.

### 3.1 Research Site

The empirical site for our research is the digital labor platform Upwork. We chose Upwork for two reasons: the extensive size of its user base and the diverse demographic profiles. First, Upwork is the largest digital labor platform, with approximately 8,000,000 freelancers and 152,000 clients [83,130]. It announced a 67% growth rate in gross revenue from freelancers and clients over the last three years [139]. This growth demonstrates that Upwork embodies a growing trend of freelance labor market participation among workers. Second, the platform provides us an opportunity to recruit workers from diverse backgrounds. Upwork's user base comprises workers from diverse gender, race, ethnicity, educational background and occupation, allowing for a more nuanced understanding of career success.

### 3.2 Data Collection

The empirical foundation is a multiyear longitudinal study involving 108 Upwork freelancers. The longitudinal design allows us to observe the nature and change of career success that emerged from workers' lived experiences as they navigate work, career and life. The data reported here were provided by freelancers who have participated in our panel study that has run once each year from 2019 to 2023, with an average interval of 12 months.

*3.2.1 Participant Sampling.* We used purposive sampling to form our panel, guided by the existing platform work literature. Probabilistic sampling is impractical due to the limited availability of complete user population data from the platforms [125,148]. Instead, we built our sample to be as diverse as possible (see Table 2).

Our criteria for selection include: workers that had earned at least $1000, doing work for United-States-based employers that we could broadly classify as administrative, technical or creative, and are active on Upwork at the time of sampling. The reasons we used broad occupational categories is because Upwork changes its occupational categories. Of course, Upwork does not provide the worker's age, gender, race/ethnicity or marital status. So, we had to invite many possible participants to the job to ensure we developed a diverse sample. We used a random number generator to drive our selection of Upwork profiles presented for each of our searches (as most searches returned thousands of possible people to hire). In doing all of this we were careful to abide by the platform's fair use policies.

Following approved research protocols, we contacted each worker and asked if they would like to be hired for a one-hour job. We paid everyone who participated in our study, ensuring that all participants received full compensation, whether they partially or fully engaged in the research. We also rated them as five stars and left strong positive reviews regarding their participation.

Over a four-year period of following the sample panel of 108 Upwork freelancers, we observed infrequent instances where participants cited demographic changes (e.g., changes in occupation or marital status) as directly influencing their views on career success during interviews. While these instances were intriguing, they were unique cases that deviated from the predominant patterns identified in our findings. Therefore, we report the initial demographic information collected during the first round in 2019.

[1] See Upwork.com

Table 2. Participant Demographics

| Category | | Total 108 | |
|---|---|---|---|
| | | Number | Percentage |
| Gender | Female | 63 | 58% |
| | Male | 42 | 39% |
| | Unknown | 3 | 3% |
| Race | Asian | 9 | 8% |
| | Black/African American | 22 | 20% |
| | White | 56 | 52% |
| | Multi-racial | 13 | 12% |
| | Other Race | 6 | 6% |
| | Unknown | 2 | 2% |
| Ethnicity | Hispanic/Latinx | 12 | 11% |
| | Non-Hispanic/Latinx | 93 | 86% |
| | Unknown | 3 | 3% |
| Location | United States | 94 | 87% |
| | Other (e.g., countries in Africa, Asia, Europe)[2] | 11 | 10% |
| | Unknown | 3 | 3% |
| Education | Post-graduate degree | 38 | 35% |
| | Bachelor's degree | 46 | 43% |
| | Associates | 7 | 6% |
| | No college degree | 15 | 14% |
| | Unknown | 7 | 6% |
| Marital status | Divorced | 5 | 5% |
| | Married | 64 | 59% |
| | Never married | 35 | 32% |
| | Separated | 2 | 2% |
| | Widowed | 1 | 1% |
| | Unknown | 1 | 1% |
| Age | 18-29 | 23 | 21% |
| | 30-49 | 70 | 65% |
| | 50-64 | 13 | 12% |
| | 65 and older | 1 | 1% |
| | Unknown | 1 | 1% |
| Occupation | Administrative (e.g., virtual assistants, bookkeepers) | 44 | 41% |
| | Technology (e.g., software developers, data scientists) | 41 | 38% |
| | Creative (e.g., graphic designers, creative writers) | 23 | 21% |
| Primary wage earner[3] | Yes | 48 | 44% |
| | No | 60 | 56% |
| Hourly wage on Upwork[4] | Median | $41 | N/A |
| | Average | $43 | N/A |

[2] The figures show participants who mentioned having worked outside the US at least once during the data collection period. These individuals maintained a work arrangement that involved traveling between the US and other countries, balancing work, vacation, and other personal responsibilities.

[3] The term indicates the individual responsible for the largest portion of the household's income.

[4] The figures are calculated based on data collected four times, with an average interval of 12 months, between 2019 and 2023.

*3.2.2 Multi-method Approach.* We employ a multi-method approach to complement existing career success research methods [120]. We pursued an explicitly multi-method approach to data collection. This makes sense in this situation as the different approaches to data collection that we use are complementary. In pursuing this, we know that career success is typically studied via archival data and self-reported data [121]. We chose to use semi-structured interviews (and a companion survey) and complement this data with archival data drawn from workers' platform profiles (with their permission).

This resulted in developing three datasets during each round of data collection: interviews, surveys and profile archival data. Each freelancer participated in a 45-minute interview. Interviews centered on two core themes: (1) how workers define and interpret success in the context of platform work, both in the short term and for their long-term careers and (2) how various technological, personal, social, and other circumstantial factors influence their interpretation of success. The interviews also explored changes in both professional and personal circumstances, including factors such as family status, employment situation, and specific occupational details.

The reality of a longitudinal study is evidenced by the variations of engagement for each participant. Out of the 108 total participants, 100 took part in interviews, which constitute the primary empirical basis of our data analysis. Among these 100 interviewees, 21 participated once, 25 participated twice, 21 participated three times, and 33 participated four times, resulting in a total of 266 individual interviews during our data collection period.

Designed as a companion to the interview, each freelancer also completed a 15-minute questionnaire. Survey questions covered complementary information to the interview such as demographic data. Lastly, we reviewed data from each participant's public platform profile, which shows their job success scores, badges earned, skills and qualifications, project histories and client reviews.

### 3.3 Data Analysis

Data analysis relied on a hybrid approach that combined inductive and deductive coding and theme development [50]. Analysis included both a deductive approach that drew on an a priori template of codes [38] and a data-driven inductive approach [24]. This was done to ensure logical consistency in researchers' interpretation of data obtained through multiple techniques across various points in time. Given the diverse properties of the dataset and the longitudinal data collection method, the analysis consisted of the following steps:

*Developing the initial coding manual:* The first author created the initial coding manual based on Seibert and colleagues' (2024) original framework [121], as presented in Table 1.

*Testing the reliability of the codes:* The first author then tested the manual's applicability by coding 16 interview transcripts, representing 15% of the research participants. The test coding uncovered additional potential dimensions not included in Seibert and colleagues' framework, necessitating a revision of the initial coding manual. The first and second authors independently coded another identical set of 16 interview transcripts, following the initial coding manual. The two authors documented additional codes and specific instances that were not covered by the initial coding manual.

*Finalizing the coding manual:* Regular weekly meetings with all authors were conducted to organize the initial codes identified from the first two stages into higher-level codes. Simultaneously, the authors conducted literature reviews to provide a conceptual foundation for the additional codes. As a result, the final coding manual encompasses twelve codes reflecting the subjective success dimensions presented in Table 1 and three codes representing the platform-mediated success dimensions detailed in Table 4.

*Coding the entire dataset:* The first and second authors systematically coded the entire dataset using the finalized coding manual. Survey and profile data were analyzed to complement the interpretation of interview data. NVivo 14[5], a computational data analysis tool, was used for the analysis of interview and survey data, while profile data was manually analyzed by extracting metadata from the archived files and integrating it with the interview and survey data sets.

*Identifying themes:* The first author grouped the codes from the fourth stage into high-level themes based on their affinity and saturated patterns.

*Corroborating and legitimizing coded themes:* The high-level themes were reviewed to ensure they accurately represented the recurring patterns. This was done through weekly meetings where all authors engaged in discussions about points of disagreement and alternative interpretations. This iterative process continued until all authors agreed on the final themes presented in this paper.

*Synthesizing themes into narratives:* The final stage involved synthesizing the themes into coherent narratives to address the research questions. The first and second authors combined the stabilized themes into narratives to provide a comprehensive presentation of the themes. The initial draft of the synthesized themes was shared with the rest of the authors, and all authors participated in reviewing the narratives.

## 4 FINDINGS

We present our findings around three insights. First, we identify seven dimensions of subjective success indicators that digital labor platform workers use to interpret their career success in platform-mediated work and careers. Second, we delineate three dimensions in which platform indicators mediate these workers' interpretations. Third, we identify two patterns in which these success indicators have evolved as workers initiate and progress in their careers on digital labor platforms. These findings provide qualitative evidence of how freelancers interpret success on labor platforms and the dimensions and indicators that are most relevant to this subset of workers.

### 4.1 Seven Dimensions of Subjective Career Success in Platform Work (RQ1)

This section presents findings addressing RQ1: How do digital platform workers interpret career success?. As we interrogated the experiences of our respondents through the lens of the 12-dimensional subjective career success that we hypothesized, we identified seven dimensions of subjective career success used by platform workers on Upwork. These seven dimensions are: (1) financial rewards (29% occurrence in the entire coded utterance), (2) agency (21%), (3) interpersonal relationship (12%), (4) fulfillment (12%), (5) self-development (8%), (6) recognition (7%) and (7) wellness (6%). Table 3 provides a summary of the seven dimensions and pertinent indicators

Table 3. Seven dimensions of subjective career success

| Dimensions | Pertinent indicators |
|---|---|
| Financial | · Decent income for achieving a basic living condition and freelancing lifestyle |
| Agential | · Control over work type, work hours and location |
| Interpersonal | · More time for familial bonds and relationships |
| Fulfillment | · Enjoyment and happiness from work |
| Self-development | · Growth in professional knowledge and skills |
| Recognition | · Validation and appreciation from the professional field and clients |
| Wellness | · Physical and emotional well-being |

[5] https://lumivero.com/resources/nvivo/

*4.1.1 Financial: "Decent income for my basic needs and freelance lifestyle".* Financial rewards are one of the most salient themes relative to freelancers' understanding of success. However, data makes clear that merely having a certain amount of money does not explain success for freelancers; instead, the values they derive from utilizing their earnings define their ideals of success. A larger number of freelancers define career success as making "just enough", "fair", or "decent" income (P10, P17, P18, P35, P39, P40, P52, P54, P57, P58, P61, P64, P65, P76, P78, P87). Among these freelancers, there is a consistent tendency for their expressions to be accompanied by projections of what they can or will do with their income. This means that freelancers value the utility of the pay, rather than the superior financial status that money might reflect. The primary utility of money emerged as meeting basic financial responsibilities such as paying rent and having additional savings to sustain freelancer lifestyles.

> "I don't necessarily need to be wealthy [...] I would define success as being able to sustain a living freelancing. It doesn't necessarily have to earn a lot of money, just able to pay my rent." — P57

> "Being able to maintain my bills and whatever else I want to be doing with my free time and extra money or savings. As long as I can maintain what I need to be doing and hopefully save a little extra, I deem myself successful." — P35

While freelancers' perspective on success extends beyond the pursuit of financial gain as a sole indicator, our data also show that earning more income is an integral part of freelancing career success. Freelancers see securing higher-paying projects and attracting more clients on freelancing platforms as key indicators of their career accomplishments (P19, P43, P45, P61, P78, P89). However, even among those who consider financial rewards as a singular criterion for success, they do not adhere to an absolute success measure such as "a certain dollar amount salary" (P89). These freelancers have a relational perspective on their income, defining their financial success compared to "the previous year" (P45) or " the last few years" (P61).

*4.1.2 Agential: "Freedom to choose where I work, how I work, the type of work I'm doing".* Having and exercising control over which work freelancers take on, where they work, when they work and how they work is another common theme in our data. Freelancers who prioritize this agential dimension articulate their success in terms of "freedom," "independence" and "autonomy" (P8, P42, P49, P67, P78, P80). From their perspectives, these descriptors form the foundation of their ability to make decisions and engage in purposeful actions when determining multiple dimensions of work such as locations, schedules, work types and modes of conducting work. And freelancers often encapsulate being able to control such dimensions as having "flexibility" (P33, P42, P49, P80, P88, P85, P91, P94). For these freelancers, career success is defined by the flexibility of time and space, enabling them to work with adaptable schedules and choose projects that resonate with their personal interests and values.

> "Flexibility, basically, is the main word. Being able to complete projects as you can or when you're willing to or able to, of course meeting your timeline of you and your clients, as well, but being able to have the freedom to basically turn on and turn off the computer or work station and work mindset as you need." — P43

> "I think it looks like having the independence and control over your time to use your time to do what you want to do that you enjoy. I think success looks like handling projects and representing clients that deal with issues that are important to you and that are enjoyable and fun and that allow you flexibility and freedom." — P49

Our data highlights that freelancers' idea of success is closely connected with the idea of 'being my own boss' through exercising agential control over the type of work they undertake. Instead of working under someone else's control, freelancers' ideals of success are linked to having a specific 'platformic status'—a status that empowers them to decline work that does not align with their needs without jeopardizing platform assets, such as client connections.

> "I think [success] would look like having opportunities come forward and knowing that I could comfortably say no, without feeling like I'm going to lose this connection or I'm going to lose that contact, or this person is never going to talk to me again, if I don't say yes to every single thing. In the most ideal situation, it would be a status where I'm able to take on projects that I really like, and I'm able to turn down something." — P104

*4.1.3 Interpersonal: "More time with my family".* Many freelancers define success as being able to spend more time with their families while effectively managing their freelancing careers (P19, P29, P48, P52, P54, P61, P76, P86, P87). Freelancers emphasize the importance of nurturing familial bonds and relationships through increased time spent together as a central marker of success. Spending quality time with children emerges as a common theme, with participants expressing joy and gratitude for the opportunity to witness their children's growth and actively participate in their lives (P19, P29, P54). The ability to be present during family dinners every night (P52) and to be with families for vacations (P48) is identified as a source of finding meaning in their success.

> "I get to be with my family every night, like I get to eat dinner with them. That's important, to be with my big family of seven, all of us together at the end of the day. [...] Did I make enough money to provide? Yes. Are my wife and are my children happy? Yes. That's what matters to me." — P52

Participants consistently weave time flexibility and financial accomplishments into their appreciation for family moments. Notably, freelancers redefine financial success by earning enough income to financially support their families (P52, P54, P61). Others underscore the significance of temporal flexibility as a key element in their career success, enabling them to be present for their children (P29, P54).

> "Success-wise I would say being able to spend so much more time with [my child], and bonding with him, and seeing him grow up. Now he's sixteen months, so that's been really just a blessing and really awesome." — P29

*4.1.4 Fulfillment: "Doing what I enjoy".* Freelancers view success as closely linked to the sense of fulfillment derived from engaging in work aligned with their passions and interests. The sense of fulfillment is manifested through expressions about their psychological state toward what constitutes a "successful career." For them, a successful career is characterized by the ability to pursue work that they find enjoyable (P12, P58, P61), that they love (P76), that brings them happiness (P90), and that adds an element of fun to their professional endeavors (P54). Freelancers use this sense of fulfillment as an inner guide, not only to assess their satisfaction with their current career but also to make decisions when contemplating about taking a different professional path.

> "Career success for me is me being happy with where I am. I'm kind of monitoring how my growth is and if this is the path that I want. [...] I'm gauging, am I happy with what I'm doing? How am I feeling with this? I can be bringing in money, but do I enjoy this?" — P45

For some freelancers, the concept of career success resides in the realm of personal happiness and contentment as they navigate career journeys on digital labor platforms (P12, P58, P65, P76).

> “I don’t want it to just be a thing of survival, I want it to be a thing of that I actually love what I do, and so far, right now.” — P76

*4.1.5 Self-development: “Being on top of my game, knowing my area, to be always learning”.* Freelancers emphasize professional growth through continuous learning to build skills and knowledge, with the ultimate goal of achieving mastery within their respective domains. Freelancers view opportunities to develop their knowledge and skill thereby being able to be the top in their field as a representation of a thriving career.

Freelancers value self-motivation and self-directed learning, as exemplified by a freelancer who expresses an unyielding desire to give their best every day. This dedication propels them into a routine of daily study, fostering a continuous evolution of professional self. This proactive approach to personal and professional development becomes a driving force in their search for work performance and excellence (P89, P67, P77, P95).

> “I want to give the best of myself every day. That really impulses me to study on a daily basis. I’m learning new things and new skills. I think that I am also becoming a leader in pushing for change and asking ourselves: how can we evolve, and be better, and perform better, and be more efficient?” — P95

Freelancers actively leverage digital labor platforms to cultivate their expertise, recognizing the pivotal role of knowledge and skills in advancing within their fields. This pursuit of mastery, often synonymous with “being at the top” of their field, becomes a salient factor for freelancer definition of long-term career success (P34, P61, P89)

> “I don’t know what I’m going to be when I grow up, at 32. I think I can say I am a this, I do this, and I’m doing it to a degree that I know I’m top in my field. I think that’s a mark of success for me.” — P34
>
> “I would say career success, I’m really big on knowledge, so for me career success is, I would say, a combination of mastery within a particular domain.” — P89

*4.1.6 Recognition: “Being respected and recognized”.* Data shows that freelancers value recognition, reputation and fame as integral components of career success. Freelancers tend to define success with reference to more concrete and tangible standards, such as publishing scholarly work in reputable academic outlets (P49), achieving "best-selling" status for book sales (P76), or gaining international recognition such as being "nominated for an Emmy" (P32).

> “I think a successful career involves scholarly articles and books that are published in great academic journals or academic presses, students that are successful in their own professional lives, and a following of people who read and are interested in your work.” — P49
>
> “So this is a little more subjective. I think career success for me would be to actually branch off and take that leap and do something for myself, even if at first when I do something for myself maybe it won’t go as successfully planned as I thought, but just piece by piece I’ll eventually get there to maybe be a best-selling writer on a certain topic or something.” — P76

On the other hand, some freelancers find success in more immediate validation and appreciation from the clients they work with. For example, P58 states that a successful career is "having clients that appreciate the work," while P86 views career success as "working with a company or clients that value both me and my work."

#### *4.1.7 Wellness: "Taking care of myself physically and mentally".* 
Holistic well-being, including physical and mental wellness, is also a dimension of freelancers' career success. Freelancers described various conditions that contribute to physical and mental health, as well as their overall definition of career success. For example, according to one participant, access to "basic human rights like health care and dental care" are conditions of having a successful career (P41). Another noted having work flexibility to take care of physical and mental well-being was important to their perception of career success (P2).

> "The flexibility to take care of myself physically and mentally. When I was practicing law the traditional way, it was just work 24/7, and I was just mentally and physically drained. Now I'm in shape. Running and exercising is part of my day because I have a flexible schedule." — P2

Conversely, some freelancers understand success through the state of emotional illbeing as opposed to wellbeing (P1, P8, P46, P52). P52 and P1 reported stress management as integral components of career success. The absence of stress is emphasized as a positive indicator, with P46 explicitly stating career success is:

> "Not having too much stress, detrimental aspects [...] I've seen how a lot of people become eroded, and jaded, and bitter towards a lot of things. They let the whole work game affect the whole dimensions of who they are as people." — P46

### 4.2 Platform Indicators: Credibility, Visibility, Predictability (RQ2)

This section reports findings answering RQ2: How do digital labor platforms shape these workers' interpretation of career success?. We find that freelancers' experiences of success are mediated by platform indicators, which include platform-generated metrics such as Upwork's job success scores (JSS[6]), client ratings, client reviews, talent badges, and additional job history insights. Data reveal that platform indicators mediate freelancers' subjective experience of success in three related dimensions: credibility, visibility, and predictability (see Table 4). Participants experience the credibility dimension of success when platform indicators enable clients to ascertain participants' reliability and trustworthiness. Visibility entails the ability for platform indicators to rank freelancers relative to each other, sifting some to the top and others to the bottom. Lastly, predictability involves the affordance that platform indicators have to enable clients to gauge how their engagement with freelancers might go, making it easier for them to make a hiring decision. The category assigned to each platform indicator is based on freelancers' interpretation about these markers and how they mediate success. Importantly, each of these dimensions results in freelancers receiving more offers or acquiring more clients.

Table 4. Three dimensions of platform indicators

| Dimensions | Pertinent indicators |
|---|---|
| Credibility | · Client ratings, client reviews, total hours |
| Visibility | · Talent badges, availability badge, diversity badges, skill certifications |
| Predictability | · Job success score, job history insights, total earnings |

[6] Upwork's Job Success Score (JSS): https://support.upwork.com/hc/en-us/articles/211068358-Job-Success-Score

*4.2.1 Credibility.* Participants report that platform indicators such as the JSS showcase their credibility to clients, making them more likely to receive more job offers. They reason that the positive indicators reflect their effort to satisfy clients. Furthermore, these platform indicators are determined by the platform, and in some cases past clients, without freelancers' direct involvement. As a result, freelancers and clients view these as an "objective" means to verify freelancers' dedication and competence. With positive indicators, participants observed that they were perceived to be credible, which in turn translated to more job offers.

> "Well, that it's managed by an independent third party and it's filtered through that third party. I think that gives people some comfort in knowing that it's not just me saying something that's true or not true. The Upwork profile system actually checks where I am, like my location; they keep track of the number of hours that I've worked through them; they highlight that I've worked a lot of hours through them and that my clients are satisfied. I think all of those independent review-type indicators are a huge selling plus for me, because they're basically vetting me without me being involved at all." – P8
>
> "[...] Having client reviews. I think just showing the credibility in whatever way that you can of the freelancer helps you get jobs." – P09

*4.2.2 Visibility.* In addition to presenting freelancers as credible, platform indicators also served as a means for potential clients to compare and sort freelancers. In turn, this sorting and filtering enabled some freelancers to rise to the top, while others remained at the bottom. In other words, some freelancers were more visible than others. For freelancers, this visibility translates to more clients reaching out to potentially hire them.

> "I think the badges helped a lot when they started giving out the Top Rated badges and things of that nature, because it lets people separate, when they're sifting through 100 people, it lets them know these are the cream of the crop off the top, and then they'll narrow it down a lot faster, and then get to me a lot faster if I've earned that badge." – P57
>
> "Well, definitely the rebuilds for my clients and the fact that Upwork is giving you different badges if you achieve different goals. I think that's super useful, because the clients will see that you have good reviews and that you have badges, and they will choose you over other freelancers" – P80

An additional aspect of visibility entails being able to access special services. For example, P37 describes receiving offers from UpWork talent people. Intriguingly, P37 receives more offers than she can handle and ends up rejecting 90% of offers.

> "As a result of getting all those badges and a high job success score, I now have access to the Upwork talent people sending me positions. That's helpful... I only accept 1 of about every 10 offers." – P37
>
> But as far as success, Upwork actually launched another tier of recognition for Upwork contractors. Before, I was just a Top Rated freelancer. Now I'm a Top Rated Plus freelancer, so that's probably the success I can think of so far. Even if I have fewer hours than what I had in the past, I'm glad that Upwork actually still recognized me among, I think it's, the top 3% in the platform." – P12

*4.2.3 Predictability.* Amidst a giant pool of credible and visible freelancers, participants note that platform indicators mediate their success by facilitating clients' ability to predict how working with them might unfold. More precisely, platform indicators such as client reviews, badges, and job success score show an overall track record of meeting past clients' expectations. As such, these indicators enable prospective clients to predict and gauge how their engagement with freelancers might go. In turn, this ability to predict makes it easier for clients to hire freelancers.

> "I think the entire experience in Upwork, because once a client sees that you already have a lot of job history, enough projects where they can see you have worked with a lot of clients…for me, there's still a difference when you're hiring somebody who's going to work in an office versus somebody who's going to be working remotely for you, so if you see somebody who has been working remotely for a number of years, you know that they can work independently, that you can trust them, that they will perform the job." – P12

> "I think the most helpful would be my skills and my past work history. Especially if the clients can see how high quality my samples are, and if they see how long I've been practicing my skills. I think there's a thing on Upwork that says, oh, you've had five years of experience with search engine optimization or three years of experience with Microsoft Word. If they see all of my different skills with various software and how I've been able to generate the work, then they'll be like, OK, then she will be great for this project." – P76

Platform indicators often interact to mediate freelancers' experience of success. As P52 notes below, the talent badges and job success score are "nice key pieces" that shape their success on the platform. Importantly, the platform indicators work together to present freelancers as credible individuals, highly-ranked workers relative to other freelancers, and promising freelancers in light of their past achievements.

> "I like some of the other pieces, like the Talent Cloud that they added. That shows everybody that I am a part of Indeed, so that gives me a little higher-level piece. It shows my education on there. It shows I'm verified, so I exist; I'm real. I like the fact that I can hide my earnings, because I've made it to that point. And I love the badges. I have a 100% job success score and I'm Top Rated Plus, so that means I provide a great experience and a great product for the user, for the client. So those key pieces are nice." – P52

### 4.3 Changes in Career Success Perspectives (RQ3)

This section presents findings regarding RQ3: How do these workers' interpretation of career success evolve over time?. Our analysis uncovers freelancers' changing perspectives on career success as they transition to and pursue work on digital labor platforms. Two discernible patterns characterize this change.

*4.3.1 From organization-centric to personal value-driven success.* First, there is a shift from emphasizing organizational success metrics to adopting a more personalized and value-driven outlook, placing importance on 'a balanced life,' manifested through a different approach to income, a desire for less stressful work experiences, more familial interactions, and attention to well-being. This shift is particularly evident among workers transitioning from traditional work arrangements to careers on digital labor platforms (P1, P35, P39, P52, P42, P45). The initial mindset, centered around organizational success metrics such as pursuing upward career mobility in organizational hierarchy, maintaining good relationships with superiors at work, and achieving financial success, has transformed into a more personal and value-driven approach. Freelancers now gauge success based on personal fulfillment, happiness and a realignment of professional values.

For example, P1 transitioned into platform-mediated work, seeking a more balanced lifestyle. Utilizing the skill sets gained from her five-year career as a researcher for a search engine company, she has accumulated 30,000 completed work hours and built relationships with two long-term clients on Upwork. While enjoying a stable career within the platform, P1 remains vigilant about her internal well-being, frequently assessing her satisfaction with both her contractual arrangements and the nature of her work. P1 reports a shift in her definition of career success.

> "I used to think of success as having climbed that top ladder in the organizational chart in my company, but now it's more of a balanced life, less stress for the job that I'm doing; being happy." — P1

Likewise, P52 ventured into freelancing after working as a manager for a major retail chain, seeking a more flexible work arrangement to balance both personal and professional life. With over five years on the freelancing platform and approximately 800 hours of work, P52 shared a profound shift in his perspective on career success. Previously driven by external, measurable financial rewards, his focus has evolved towards intrinsic and subjective elements, particularly the freedom to structure his work around his life with family.

> "The freedom. I have five kids, I'm married, and it really gave us the freedom to balance and make my own hours around my life. Because I always chased my career, like I always kind of lived to work, and now I work to live, really. It's more that I switched and have a different outlook on life. At the end of the day, I'd always say how would I gauge my success, financially? Did I make my boss happy? Now it's changed. I've had so much change in even my own personal values, where I was always looking inwardly for success, meaning I was striving for that financial success, but also, on the other hand, you lose a lot in life. And we only get one life." — P52

*4.3.2 From securing job for financial rewards to finding quality work for self-growth.* Freelancers express a shift in their views from seeking jobs for financial gain to actively pursuing quality work for professional growth. This pattern is observed among workers initially driven by a financial necessity to work online. However, as they continue on digital labor platforms, their definition of success expands to encompass the pursuit of professional growth through quality job opportunities provided by the platform.

For example, P15, an editor on Upwork with a background in the publishing industry and postgraduate training was initially driven to work online in search of income. However, she shared how her definition of success has shifted from achieving financial rewards to being engaged in quality work in order to grow her career and eventually produce written works under her own name. At the beginning of her freelance career in 2020, her work strategy revolved around securing job opportunities, as she states, "Whatever editing jobs are posted, I will put bids in on the ones I think fit into my skill set." And she evaluated her success by the monetary compensation received. After three years, she shared that her views had shifted from simply getting work and getting paid to selectively accepting jobs that provide her with "skill expanding opportunity" and "good use of time for the compensation".

Similarly, P53 initially had a professional career within a conventional organization, serving as a senior administrative staff member. The transition to online freelancing commenced out of financial necessity, he viewed "getting any job" as success when he embarked on this journey in 2020. Contrastingly, by 2023, not only had his concept of success expanded, but so has his career on the platform. Establishing a solid 'platformic status' empowered him to selectively decline projects that he deems unworthy of his time and dedication.

> "Yes, my definition has changed because my expectations have probably changed. It's had to grow as I expect more out of it. Before, I was just grateful to have anything and now I realize I can, because of my skills and I do have the clientele, I can say no to some jobs and I can take on others." — P53

## 5 DISCUSSION

The findings reported here demonstrate how workers' interpretations of career success are evolving and how digital work platforms are shaping this transformation. Indeed, the empirical focus on online freelancing provides a direct contrast to traditional conceptualizations of career success. Building from this, we contribute to current conceptualizations of career success in platform-mediated work and careers, connecting the findings, concepts and insights from the careers literature with that of CSCW. In this discussion, we focus on the roles of platform metrics as new success indicators shaping workers' experiences on digital platforms. We also highlight the precarious realities of platform-mediated careers with the effort to find work and the uses and designs of the platform bound up together in a sociotechnical ensemble [77,88].

### 5.1 Platform indicators

Our findings show that platform indicators allow workers to show their credibility, gain more visibility, and offer clients assurances of predictability. However, the standardized, platform-driven and opaque nature of these indicators also diminishes workers' control over their success, as their career narratives are controlled by the platform's and subject to the platform's interests. In essence, while success ideals may originate from workers' intrinsic values, their realization is often linked to the external control exercised by digital labor platforms.

*5.1.1 Paradoxical affordance affecting success in platform-mediated careers.* Findings show that platform indicators function as de facto "objective" success indicators. The three traits of objective success are being directly observable, measurable, and verifiable by third parties [67,71,72]. Platform-specific metrics (such as five-star ratings, job success scores in percentiles, and tiered badges) collectively constitute an "objective" template that documents and quantifies workers' project history skills, and client satisfaction on the platform. Moreover, freelancers view indicators as managed by an "independent third party" (P8, P39, P95). This perceived impartiality and objectivity form workers' trust toward the platform indicators. Freelancers must rely on platform indicators, as there is no alternative "objective" indicator unless they collect their "success evidence" from actors involved in their platform work, such as clients (P1, P8, P15, P57, P95).

Beyond acting as objective measures, platform indicators on Upwork confer status and titles, such as "Top Rated" freelancer (signifying among the top 10% of freelancers within the labor market hierarchy). This status implies greater credibility and visibility, resulting in more client attention and ultimately, more success on the work platform.

Ashford and colleagues (2018) questioned, "What metrics will workers use to define success when the traditional career ladder with its promotions and salary increases becomes less relevant?" [13:25]. Our research presents one possible answer to this question: Platform-driven indicators replace the traditional promotion ladder. In the case of online freelancers, a tiered badge system that stratifies workers within the platform labor market. This presents freelancers with corresponding job opportunities and a further avenue for achieving accompanying success. The salary in traditional careers is replaced by workers' pay rate in platform work, which can increase over time to enable financial success. Thus, we suggest that platform indicators take the role of objective success indicator by which workers gauge their past, current and predicting their future success on the platform.

While platform indicators provide workers an avenue to achieve career success, their opaque nature may hinder workers' success experiences. Our findings demonstrate that freelancers' career success is essentially constructed through their “logged labor” [141] under the purview of “platformic management” [79]. Freelancers' interactions with clients, hours invested in projects, and the application of skills and knowledge translate into standardized platform indicators that are calculated by ostensibly “objective” algorithms. Yet, existing research demonstrates that algorithms reflect platforms’ own interests and machine learning algorithms that calculate metrics mirror the explicit or unconscious human bias and mistakes prevalent in the traditional labor market [2,3,15,89,97,110]. Consequently, these indicators may autonomously impact workers' experiences, including performance evaluation [43,78,95], search ranking visibility [32,41,147], and future client attention and hiring decision [39,114].

The notion that career success is demarcated by platform indicators suggests that the designed platform ecosystem becomes a clear avenue for long-term platform solvency. Despite the optimism of “an emerging networked society of microentrepreneurs” [135:176], the reality is career success is intimately tied to specific platform ecosystems in two ways. First, freelancers are required to consent to platform rules and policies thus surrendering their career success to platform-managed metrics. Second, indicators are platform-specific, meaning that career success becomes directly tied to an individual platform. The result is that the “portability” of career success is limited, and over time further indentures career success into the platform ecosystem.

Beyond mediating job-level success, platform indicators also shape how workers view their career success. Within this 'platform-mediated success narrative,' workers may not have much control over their careers to align with their subjective ideals of success. Instead, platform metrics serve as a prerequisite for workers' platformic status, playing a pivotal role in determining their ability to achieve success ideals such as income, spatio-temporal agency, and self-development.

In light of the paradoxical nature of platform indicators, we call for greater attention to problematize the construction and influence of platform indicators on workers' success, both at the job and career levels. Existing literature from CSCW and cognate communities on algorithmic management documented ample evidence of the political, unethical and partial nature of algorithms embedded in digital labor platforms [15,28,66,85,104,117]. Future research critically examines the formation and operation of platform-driven success indicators, as well as their impact on the short-term and long-term success of digital platform workers. By doing so, we can strengthen the positive impact of objective platform success indicators while mitigating any adverse effects they may have.

*5.1.2 Standardization of success: reducing the rich tapestry of success to a predefined template.* Another issue related to platform indicators is their role in offering a standardized template against which the success of workers' careers is judged and might be deemed insufficient. Given our findings indicating that workers' career success interpretation involves multiple dimensions, we call for critical scrutiny into the extent to which existing platform indicators may undermine such diverse notions of success. Indeed, freelancers view success as a holistic concept: "a complete package" encompassing various aspects of work and life, such as time flexibility, a satisfactory salary and quality time with family. This suggests that the seven subjective career success dimensions identified here may not be mutually exclusive but often coexist and complement each other.

Nevertheless, platforms offer predetermined templates that deconstruct the holistic values of workers' success and selectively emphasize aspects pertinent to platform transactions, such as skills and financial gains [102]. These platform design choices promote institutionalized ideals of career success, creating a benchmark against which workers compare themselves and engage in competition. Building on [60:126], [41] argues that platforms create a hierarchical space in which “all participants relate to one another continuously and competitively.” The standardization of success on digital work platforms may constrain the variety of success stories and the directions workers wish to carve out in their careers. This might, then, impede the evolution of worker-centered definitions of success within this career trajectory, making freelancers subordinate to what platforms define as "successful.”.

Future research should critically examine how platforms contribute to the construction of institutionalized ideals of career success, potentially limiting the diversity of success narratives. To this end, two inquiries are worth exploring. First, researchers can delve into the impact of the three platform indicators on influencing the extent to which digital platform workers experience the seven subjective career success dimensions. Existing career research reveals the interplay between objective career success and subjective career success [1,12,121,123,131]. Given the objective nature of the platform indicators discussed, understanding how these dual facets of career success interact can illuminate the potential impact of platforms on workers' interpretation of career success. Second, on the platform side, researchers should also explore strategies and interventions that digital labor platforms could employ to embrace a more inclusive and diverse evaluation and presentation of digital platform workers' success. This investigation would help accommodate the diverse values and aspirations highlighted in our findings within the design and operation of digital labor platforms.

## 5.2 Platform-mediated Careers

Findings make clear that success in freelance work reflects a more individually-centered and subjective perspective. This differs from the more well-known, and perhaps default, views of career success as organizationally-centered and tied to financial gains and increased responsibilities. This success, however, is mediated by the platform where there is no assurance of either financial gain or increased responsibilities.

*5.2.1 Barriers to achieving career success.* Our findings highlight the barriers associated with pursuing career success in the context of platform-mediated careers. Framed by the precarious nature of non-standard work (e.g. freelance work), platform-mediated careers face at least two additional barriers to achieving career success. First, and like all gig-based workers, online freelancers face a litany of additional barriers they have to navigate on the platform [66,77,79]. At the job level, freelancers express concern about the erratic flow of projects, leaving them in a state of indefinite waiting for the next project (P47, P66). Moreover, there is an absence of forecasting mechanisms to indicate that such uncertainty will stabilize. If a career consists of an ongoing series of daily work experiences, the unpredictable nature of projects signifies a potential disruption in freelancers' career trajectory. This lack of predictability can hinder freelancers from attaining sustained work fulfillment, gradual development, and more tangibly stable financial rewards, all of which are reported as some of the important success dimensions in our findings.

Second, against this backdrop of precariousness, the importance and opacity - magnified by constantly changing platform indicators (e.g. Top Rated Performer, etc.) contribute to the instability that freelancers face in their pursuit of subjective career success. As previously discussed, the opaqueness of these indicators subject freelancers to “platformic management” such that their agency in work-related decision making is dominated by the platform [79]. Furthermore, the seemingly objective nature of these platform indicators does not always align with freelancers’ definitions of career success. This discrepancy, in turn, acts as a barrier between freelancers’ pursuit and attainment of career success.

These barriers are in line with extant literature concerning the precarious nature of platform work and career [13,23,30,31,88]. Existing work discusses career path uncertainty in conjunction with identity challenges [13], career-related competency building [90] and career progression [31]. Extending such discourse, our findings suggest that career path uncertainty is one potential obstacle to attaining success that digital platform workers value. To mitigate such an obstacle, future research should explore strategies and interventions tailored to offer these workers a more transparent and defined career trajectory.

*5.2.2 Changing nature of career success.* The longitudinal data allow us to see that freelance workers' interpretation of career success evolves over the course of their careers. As individuals transition from the traditional "organization man" to "digital labor platform workers," career success shifts from conventional organization-centric perspectives to a model driven by personal values. Then, as individuals progress from being "newcomers" to becoming "experienced" on digital labor platforms, their views on career success evolve from an emphasis on immediate job acquisition and financial gains to a focus on longer-term professional growth.

This evolving nature of career success merits further attention. Existing platform work scholarship in the CSCW and related communities provides ample insight into how multidimensionality of workers influences their experiences and outcomes in platform work. Such dimensions include gender, race, and class [103,112,144], occupation [101,105], socio-economic status [14,69], and individual motivations and priorities [44]. Similarly, career studies scholars have documented extensive evidence indicating that individual, situational, and social factors significantly influence career actions and outcomes, including subjective and objective success [92,106].

Future investigations should build upon the identified success dimensions and contextualize them within the broader spectrum of workers' career circumstances and structural conditions. Such endeavors are expected to unveil the dynamic relationship between success dimensions and factors that potentially shape how workers perceive and navigate career success in platform-mediated work and careers.

## 5.3 Design Insights for CSCW

The findings and discussion lay the groundwork for exploring future design practices aimed at enhancing computer-supported career development. We propose three potential approaches to place a worker-centered perspective on success at the core of these practices.

*5.3.1 Transparency-Oriented mechanisms.* We raise concerns about the opaque and volatile nature of platform-mediated success indicators, which exacerbate the precariousness of career success on digital labor platforms. Future CSCW research and design practices could mitigate these barriers through two possible research directions.

Platform-indicator management tools: Platforms like Upwork currently provide abstract equations used to quantify success indicators (e.g., the Job Success Score). Workers need to understand these equations and to gauge if their actions on the platform favor such equations, creating another form of invisible unpaid labor. Future research can explore developing intelligent tools, such as chatbots or dashboards, that explain how platform-mediated success indicators are calculated. More importantly, these tools should offer insights on specific actions needed to improve platform indicators, helping freelancers develop strategies for clearer, more predictable long-term career success on the platforms.

Career projection/planning tools: Freelancers often experience precarious career paths marked by inconsistent job availability, financial instability, limited opportunities for ongoing skill development, and difficulties in planning long-term career development. Future research should explore design interventions to address these sources of uncertainty inherent in platform-mediated work to promote long-term career success. Potential areas of focus could include developing predictive models to forecast job availability, creating tools for planning and forecasting hourly rates based on financial goals, implementing structured skill development tools with targeted learning resources, and developing computer-supported mentorship tools to facilitate comprehensive career planning and sustained success.

*5.3.2 Diversity-Sensitive Design.* We encourage platform owners to move beyond homogenized notions of success, as this approach diminishes the range of values and meanings workers derive from their work. Future CSCW design practices should seek ways to reflect such diverse values in platform design choices. We discuss two directions.

Growth/process-centered features: The current platform indicators tend to be outcome-driven (e.g., the number of successful jobs completed, ratings given at the end of projects), which neglect freelancers who define self-development as one of important dimensions of career success. Future research can explore how platform designs can reflect the growth of skill sets and increased knowledge. This approach allows workers to emphasize the process and gradual development that they experience and appreciate as part of their career success, signaling their continuous effort in developing professional competencies to future employers on the platform.

Role diversity features: Embracing diverse life roles beyond freelancing is one of the key aspects of how workers define success in platform-mediated careers. Future research should focus on exploring ways to integrate workers' professional and personal roles in their profiles, acknowledging that these roles can coexist as success indicators. For instance, parents who prioritize spending more time with their children should be able to showcase this preference in their platform work through a "Parents Badge." Such visual displays of success ideals would allow workers to express their preferred success indicators and help negotiate clients' expectations regarding freelancers' preferred work-life boundaries focused on family.

*5.3.3 Portability-Enabling Design.* We discuss that the portability of career success is limited, further indenturing career success into the platform ecosystem as workers must follow platform-defined success rules and policies. To enable the portability of career success and mitigate these limitations, future research should explore a unified portfolio system that allows freelancers to export their work history, ratings, and client reviews from one platform to another. This feature would enable freelancers to showcase their "success portfolio" across multiple professional platforms, demonstrating their professional achievements.

Finally, we note that the proposed ideas necessitate a careful discussion of their feasibility and sustainability, as the varied priorities and interests of stakeholders within the digital labor platform ecosystem may introduce barriers or suggest alternative approaches to the practical design and implementation of these concepts.

## 5.4 Future Research

Our work lays the groundwork and surfaces insights into career success in platform work. To further advance the contributions of our paper, we propose two approaches for future research endeavors.

*5.4.1 Intersectional approaches.* The meaning of success in online freelancing could provide different insights when viewed through the lens of diverse demographic positions such as gender, ethnicity, and occupation. To gain a more nuanced understanding of career success in digital labor platforms, future research should incorporate intersectional approaches to study how overlapping identities (e.g., gender, race, class, and occupation) collectively influence freelancers' definitions and changes in career success, revealing the compounded effects of multiple social identities on career trajectories and success metrics on digital labor platforms.

*5.4.2 Cross-Platform Studies.* Platform work encompasses diverse sub-genres such as online freelancing, crowdwork, delivery, transportation, and care work [87]. This heterogeneity offers an opportunity for future CSCW research to go beyond our current empirical investigation focused on online freelancing. One promising direction would be to conduct cross-platform studies that compare career success dimensions and indicators, as well as workers' perspectives, across different sub-genre of platform work. This approach would enable researchers to identify both commonalities and unique experiences of success, providing a deeper understanding of the discourse on the meaning of success within platform work scholarship.

## 6 CONCLUSIONS

We contribute to an understanding of the meaning of success on digital labor platforms from worker-centered viewpoints. Insights presented in this research advance four contributions to the CSCW and cognate communities.

First, our study contributes three empirical insights to elucidate the meaning of success as viewed through workers' eyes, experienced through platform intermediation and changed over the course of workers' careers. We provide this contribution through the formation and empirical validation of seven subjective success indicators in platform work. Relatedly, we identify how these perceptions of success evolve over time through our longitudinal investigation. In doing so, we respond to calls to investigate the meaning, nature, and process of success in platform work.

Second, we contribute to our understanding of the roles of platforms in shaping career success. We outline three dimensions in which platform indicators mediate digital platform workers' perceptions and experiences of career success. Our work demonstrates the nature of platform indicators not only on the job-level success of digital platform workers, but also on their career success. In doing so, we contribute to research on how platformic management manifests itself in career success.

Third, the conceptual developments provide a bridge between contemporary CSCW and careers scholarship. Historically, CSCW scholarship is concerned with understanding and designing computing technologies that mediate interactions among people in work settings [34]. With the development of advanced computing technology, the concept of 'work' has evolved from a single task to work practices, and from work practices to career [119]. Our research demonstrates that the rise of digital labor platforms illustrates how computing technologies not only mediate work at the task and practice levels, but also exert influence at the career level. In doing so, we contribute to our understanding of contemporary careers as mediated by digital labor platforms.

Finally, we contribute to design and research practices by providing three design insights for creating platform work environments conducive to worker-centered success. These insights emphasize transparency, diversity, and portability to improve the overall experience and success of workers on digital platforms. We also suggest two future research directions: adopting an intersectional approach and conducting cross-platform comparisons to advance the discourse on success in platform work scholarship within the CSCW community.